\documentstyle[12pt]{article}
\begin{document}
\def\theequation{\arabic{section}.\arabic{equation}}
\newcommand{\be}{\begin{equation}}
\newcommand{\ee}{\end{equation}}
\begin{titlepage}
\title{A singularity--free cosmological model with a conformally
coupled scalar field}
\author{S. \c{S}. Bayin \\ {\small \it Department of Physics,
Middle East Technical University} \\
{\small \it Ankara (Turkey)} \\ \\F. I. Cooperstock
\\ \\ and \\ \\ V. Faraoni
 \\ \\{\small \it Department of Physics and Astronomy, University
of Victoria} \\
{\small \it P.O. Box 3055, Victoria, B.C. V8W 3P6 (Canada)}}
\date{}
\maketitle   \thispagestyle{empty}  \vspace*{1truecm}
\begin{abstract}
We explore the possibility of describing our universe with a
singularity--free, closed, spatially homogeneous and isotropic
cosmological model, using only general relativity and a suitable
equation of state which produces an inflationary era. A phase
transition to a radiation--dominated era occurs as a consequence of
boundary conditions expressing the assumption that the temperature
cannot exceed the Planck value. We find that over a broad range of
initial conditions, the predicted value of the Hubble parameter is
approximately $47$ km$\cdot$~s$^{-1}\cdot$~Mpc$^{-1}$. Inflation is
driven by a scalar field,
which must be conformally coupled to the curvature if the Einstein
equivalence principle has to be satisfied. The form of the scalar
field potential is derived, instead of being assumed a priori.
\end{abstract}
\vspace*{1truecm} \begin{center} {\small
To appear in {\em The Astrophysical Journal}.} \end{center}
\end{titlepage}   \clearpage
\section{Introduction}

Singularity--free cosmological models have received renewed attention
in the past few years (Rosen 1985, Israelit \& Rosen 1989 hereafter IR,
Balbinot {\em et al.} 1990, Starkovich \& Cooperstock 1992 hereafter SC,
Brandenberger {\em et al.} 1993). In the paper by IR, the universe was
modelled as
a closed Friedmann--Lemaitre--Robertson--Walker (FLRW) spacetime, bouncing
from a Planck mass and radius state with inflation, emerging into the
standard model of radiation
and then matter dominance. The IR ``prematter'' heated with expansion
as it emerged from the cold non--singular Planck state. This is an
attractive simple picture, in contrast to the earlier prevailing theme
among cosmologists that the universe emerged from a singular state in
the big bang into a hot radiation--dominated universe, cooled to the
point of vacuum energy dominance leading to the onset of inflation
with further cooling. Then scenarios had to be constructed to re--heat
the universe to the onset of the radiation--dominated phase of the
standard model.

SC built upon the IR model, incorporating a scalar field
to describe all phases of the universe evolution as an integrated,
albeit simplified, field theory. Besides being a physically meaningful
source of gravitation in the early universe, a scalar field provides a
perfect fluid stress--energy tensor also in the later epochs of the
universe's history. Moreover, it has been proposed as a dark matter candidate
(Starkovich 1992; Kofman {\em et al.} 1993; Delgado 1993; McDonald
1993).

We adopt the Gliner (1966)--Markov (1982) picture of
Planck limits to physical quantities and postulate that when the
Planck temperature $ T_{pl}$ is reached, the phase change from
prematter (with equation of state $P=(-1+ \gamma ) \rho $, where $\gamma $
is a small parameter) to radiation ($P=\rho /3 $)
occurs.

In this paper, the work of SC is extended, refining the
calculations of the evolution of the scale factor. We now determine
the value of the present time as that corresponding to the evolution
of the radiation temperature to 2.7~K. We then find that for a broad
range of initial conditions (no ``fine--tuning'' requirement), the
present value of the Hubble parameter is approximately $47$
km$\cdot$~s$^{-1}\cdot$~Mpc$^{-1}$.

The paper by SC is built upon the minimal coupling of
the scalar field to the Ricci curvature.  While minimal coupling
is usually considered, it has been shown recently (Sonego \& Faraoni
1993) that conformal coupling is the only acceptable possibility if
the Einstein equivalence principle holds as applied to scalar wave
propagation. Thus, it is of immediate interest to focus our cosmological
theory on a conformally coupled scalar field.

Instead of
adopting a particular form of the scalar field potential (such as, e.g.,
$\lambda \phi^4$) as is usually done, we start with the assumption that
the equation of state is
inflationary, and derive the form of the potential. This approach
reverses the procedure usually employed, but is
perfectly legitimate, and has already been used in the literature
(Lucchin \& Matarrese 1985, Barrow 1990, Ellis \& Madsen 1991, SC).
Also, to be noted is a somewhat similar approach, deriving the scalar
field potential from the observed spectrum of density perturbations (Hodges
\& Blumenthal 1990, Copeland {\em et al.} 1993, Turner 1993, Lidsey \&
Tavakol 1993).

We regard as acceptable those cosmological models that merge
satisfactorily with the standard model without singularities and
without the need for fine--tuning of the parameters. We are able
to impose the conditions which are deemed necessary and deduce the
potential which achieves these aims.

In sec.~2 we present the basic equations of the cosmological model with
a scalar field. In sec.~3 we introduce boundary and initial conditions
inspired by the Gliner--Markov ideas, and we solve analytically the
equations for the dynamics of the universe. In sec.~4 we solve
numerically for the evolution of the scalar field and we find the form
of the scalar field potential. The results are discussed in sec.~5.

\section{Basic equations}

Our theory is based on a closed, spatially homogeneous and isotropic
FLRW model. The metric is given
by\footnote{We adopt the notations and conventions of
Birrell \& Davies (1982).}\setcounter{equation}{0}
\be     \label{1}
ds^2=dt^2-a^2(t) \left[ \frac{dr^2}{1-r^2}+r^2 \left( d\theta^2
+\sin^2 \theta \, d\varphi^2 \right) \right] \; ,  \ee
where $t$ is the comoving time, $r \in (0,1)$, $\theta \in
(0,\pi)$, and $\varphi \in (0,2\pi)$. We assume that the only source
of gravitation is a scalar field $\phi(t)$ during the entire history
of the universe. In the contemporary literature it is customary to
assume that the dynamics of the universe are dominated by a scalar
field during an early inflationary epoch. We extend this assumption to
the later radiation-- and matter--dominated epochs, in which a scalar
field provides the simplest field--theoretic realization of a perfect
fluid stress--energy tensor. However, the use of the scalar field as a
source of gravitation in these late eras may be more meaningful; in
fact it has been pointed out (Starkovich 1992, Kofman {\em et al.}
1993, Delgado 1993; McDonald 1993) that
the $\phi$--field could be a candidate for dark matter in the present
era.

Most of the contemporary literature is built upon the minimal coupling of
the scalar field to the Ricci curvature $R$. The Klein--Gordon equation
with more general coupling is
\be    \label{2}
\Box \phi+\xi R \phi +\frac{dV}{d \phi}=0 \; ,
\ee
where $\Box \phi \equiv g^{\mu\nu}\nabla_{\mu}\nabla_{\nu} \phi$, which
assumes the value $\ddot{\phi}+3\dot{\phi} \,\dot{a}/a$ for our metric,
$ \xi$ is a
numerical constant, and $V( \phi)$ is the scalar field potential. While
minimal coupling ($\xi=0$) is usually considered, other values of $\xi$ are
possible. Recently, Sonego \& Faraoni (1993) showed that the
value $\xi=1/6$ (``conformal coupling'') is the only acceptable possibility
if scalar waves in a curved spacetime are to propagate {\em locally} as
they do in Minkowski spacetime, as required by the Einstein equivalence
principle.
We stress that the derivation of this result is independent of the
conformal structure of spacetime and of conformal transformations: in
other words, conformal invariance of the Klein--Gordon equation is not
required. In fact, eq.~(\ref{2}) with any non--constant potential (or
a mass term) is not conformally invariant\footnote{This can be understood
physically
by noting that the introduction of a mass, or a potential, sets a
preferred length scale for the scalar field theory, which is then not
invariant under a change of lengths and norm of vectors arising from a
(position dependent) rescaling of the metric. Thus, conformal
invariance follows from ``conformal'' coupling only in the absence of
a potential.}.

Thus, we focus our cosmological theory on a conformally coupled scalar
field. The stress--energy tensor is (see e.g. Birrell \& Davies
1982, p.~87)
\be                \label{3}
T_{\mu\nu}=\frac{2}{3} \nabla_{\mu}\phi \nabla_{\nu}\phi -\frac{1}{6}
\nabla_{\alpha}\phi \nabla^{\alpha}\phi \, g_{\mu\nu}-\frac{1}{3}\,\phi
\nabla_{\mu}\nabla_{\nu}\phi -\frac{1}{6}\,\phi^2 \left(
R_{\mu\nu}-\frac{1}{2} \, g_{\mu\nu} R \right) +V
g_{\mu\nu}+\frac{1}{3}\,\phi \,\Box \phi \, g_{\mu\nu} \; . \ee
This can be written in the form of a perfect fluid stress--energy
tensor:
\be    \label{4}
T_{\mu\nu}=\left(\rho +P \right)u_{\mu}u_{\nu}-P g_{\mu\nu} \; ,
\ee
where
\be    \label{5}
u^{\mu}=\frac{\nabla^{\mu}\phi}{\left( \nabla_{\alpha}\phi\nabla^{\alpha}
\phi \right) ^{1/2}}
\ee
is the four--velocity of comoving observers (we restrict ourselves to
the consideration of a real scalar field which satisfies
$\nabla_{\alpha}\phi\nabla^{\alpha}\phi>0$). The energy density and pressure
are given by (a dot denotes differentiation with respect to $t$)
\be    \label{6}
\rho=T_{00}=\frac{1}{2}\, ( \dot{\phi})^2+\frac{1}{2} \, \phi^2\left[
\left( \frac{\dot{a}}{a} \right)^2+\frac{1}{a^2} \right]+\frac{\dot{a}}{a}\,
\phi \, \dot{\phi}+V
\; ,   \ee
\be     \label{7}
P=\frac{1}{3} \left( \rho-{T^{\mu}}_{\mu} \right)=
\frac{1}{6} \,( \dot{\phi})^2+\frac{1}{3}\,\frac{\dot{a}}{a}\, \phi \,
\dot{\phi}+\frac{1}{3}\,\phi \,\frac{dV}{d\phi}+\frac{1}{6}\, \phi^2\left[
\left( \frac{\dot{a}}{a} \right)^2+\frac{1}{a^2} \right]-V\; ,
\ee
where eq.~(\ref{2}) has been used.
It has been shown by Madsen (1988) that the stress--energy
tensor of a scalar field can be put in the form corresponding to a
perfect fluid only under the assumptions of spatial homogeneity and
isotropy. If spatially dependent perturbations $\delta \phi (t,\vec{x} )$
to the scalar field are considered, $T_{\mu\nu}$ can still
be put in the form corresponding to a fluid, but new terms appear,
which can be interpreted as a heat flux and anisotropic stresses.
These terms disappear in the case of minimal coupling (Madsen 1988).
This has, as a consequence, the fact that, during a given era, there is
no entropy production for the unperturbed field $\phi(t)$. This
is also true for spatially dependent perturbations of a minimally
coupled scalar field, but, interestingly, it no longer holds for conformal
coupling, where entropy production does occur.
To our knowledge, perturbations of a non--minimally coupled scalar
field have not been considered in the literature. We will not deal
with this in the present paper, leaving it for future analysis.

The Einstein equations
\be     \label{8}
R_{\mu\nu}-\frac{1}{2}g_{\mu\nu} R=-8\pi \, T_{\mu\nu}
\ee
give
\be     \label{9}
\left( \frac{\dot{a}}{a}\right)^2+\frac{1}{a^2}=\frac{8\pi}{3}\, \rho \; ,
\ee
\be               \label{10}
\frac{\ddot{a}}{a}+4\pi \left( \gamma-\frac{2}{3}\right) \rho=0 \; ,
\ee
where $\gamma$ is defined by the equation of state
\be    \label{11}
P=\left( \gamma-1 \right) \rho \; .
\ee
For simplicity, we assume that $\gamma$ is constant during each
era in which we divide
the history of the universe. We will consider three eras,
corresponding to different values of $\gamma$. We consider first
the evolution of the scale factor, and then we will turn to the
dynamics of the scalar field.

\section{The dynamics of the universe}

The evolution of the scale factor can be solved separately from the
dynamics of the scalar field. The Einstein equations (\ref{9}) and
(\ref{10}) give an equation involving only the scale
factor\setcounter{equation}{0}
\be  \label{12}
\frac{\ddot{a}}{a}+\left( \frac{3}{2}\, \gamma-1 \right) \left(
\frac{\dot{a}^2+1}{a^2} \right) =0  \; .   \ee
In order to solve this equation, we introduce the conformal time $
\eta$ defined by
\be \label{trans}
dt=a( \eta) d\eta   \; .
\ee
In the following, we will solve for the dynamics of the universe and
of the scalar field in terms of the conformal time. The transformation
to comoving time is given explicitly in eq.~(\ref{52}), and the final
results are also given in comoving time.

Equations (\ref{9}), (\ref{10}) and (\ref{12}) become
\be     \label{9bis}
\left( \frac{a'}{a^2}\right)^2+\frac{1}{a^2}=\frac{8\pi}{3}\, \rho \; ,
\ee
\be               \label{10bis}
\frac{a''}{a^3}-\left( \frac{a'}{a^2}\right)^2+4\pi \left( \gamma-
\frac{2}{3}\right) \rho=0 \; ,
\ee
\be  \label{12bis}
\frac{a''}{a}+\left( \frac{3}{2}\, \gamma-2 \right)
\left( \frac{a'}{a}\right)^2+\frac{3}{2} \,\gamma-1=0  \; ,
\ee
where a prime denotes differentiation with respect to $\eta$. If we focus
our attention on the variable
\be    \label{13}
u \equiv \frac{a'}{a}=\frac{d \ln a}{d\eta} \; ,
\ee
we obtain, from eq.~(\ref{12bis}), the Riccati equation
\be \label{14}
u'+cu^2+c=0 \; ,
\ee
where
\be \label{15}
c=\frac{3}{2}\,\gamma-1    \; .
\ee
In the following, we will consider values of $\gamma$ such that $c\neq
0$. Equation (\ref{14}) is easily solved by setting
\be   \label{16}
u=\frac{1}{c} \, \frac{w'}{w}=\left[ \ln \left( w^{1/c} \right) \right]' \; ,
\ee
which gives the real solution
\be   \label{17}
a( \eta)=a_0 \left[ \cos ( \eta c +d) \right]^{1/c} \; ,
\ee
where $a_0$ and $d$ are integration constants.

We consider three periods in the history of the universe:
\begin{itemize}
\item {\em ``prematter'' (inflationary) era}, $0 \leq \eta \leq
\eta_r$, in which $\gamma$ is small and positive (typical values are $
\gamma_p \sim 10^{-3}$). This gives an inflationary equation of state
close to $ P=-\rho$ (the case $ \gamma_p=0$ corresponding to eternal
inflation is excluded). The smallness of the parameter $\gamma_p$ determines
the amount of inflation that the universe has experienced.

\item {\em radiation era}, $\eta_r \leq \eta \leq \eta_m $, with
$ \gamma_r=4/3$, corresponding to the equation of state $P=\rho/3$.

\item {\em matter era}, $\eta \geq \eta_m$, with $\gamma_m=1$,
corresponding to a zero pressure dust
\end{itemize}
(in the following, subscripts or superscripts $p$, $r$, $m$ denote the
prematter, radiation and matter era, respectively). In our model, the
transitions between different eras, which
occur at the conformal times $\eta_r$ and $\eta_m$ are imposed by
discontinuous changes in the equations of state. The phase change at
$\eta_r$ follows from the Gliner (1966)--Markov (1982) hypothesis
of Planckian limits
to physical quantities, in this case temperature, and the phase change
at $\eta_m$ occurs when the radiation density falls to the level of
the non--relativistic matter density. Although the imposition of
discontinuous phase changes is not derived from an analysis of the
microphysics, the microphysics relating to the scalar field is itself
unknown (Ellis 1991). However, the present procedure does yield
the form of the scalar field potential $V( \phi)$, in contrast to the
usual procedure of assuming, {\em a priori}, a form for $V( \phi)$. This is
a variation of the approach first proposed by Synge (1955) and
which has since been followed by various authors (Lucchin \&
Matarrese 1985, Barrow 1990, Ellis \& Madsen 1991,
SC -- see also Hodges \& Blumenthal 1990 and Copeland {\em et al.} 1993).
As applied to cosmology, this
leads to the treatment of the early universe as a ``hot laboratory for
the particle physics relevant at these very early times'' (Ellis \&
Madsen 1991).

\subsection{Boundary conditions and solutions for the scale factor}

Our initial and boundary conditions for the dynamics of the universe
follow from the idea that the physical quantities are limited by the
Planck values. This idea is present in the papers of Gliner (1966) and
Markov (1982), and has been used by Rosen (1985), IR and SC. The closed
cosmological model
has a contracting phase preceding the turnaround, or ``bounce'', when
the Planck density is reached. We set $\eta=0$ to be the time of the
bounce, at which point the scale factor has a vanishing
derivative\footnote{The possibility of imposing the initially static
condition
(\ref{18}) is by no means trivial, and is guaranteed by the
inflationary equation of state that we assume, which circumvents the
Hawking--Penrose singularity theorem. An equation of state satisfying
the strong energy condition would imply a singular derivative of the
scale factor, and an initial singularity. The inflationary equation of
state violates the strong energy condition.},
\be   \label{18}
a'(0)=0 \;,
\ee
\be   \label{19}
\rho(0)=\rho_{pl} \;,
\ee
(where $\rho_{pl}=5.1566 \cdot 10^{93}$ g~$\cdot$~cm$^{-3}$ is the
Planck density), which also imply
\be   \label{20}
a(0)=a_{0}^{(p)}=\sqrt{\frac{3}{8\pi \rho_{pl}}}  \;,
\ee
\be   \label{21}
\rho'(0)=0 \;.
\ee
At this stage, the universe is said to be in the ``prematter'' phase.
Equation (\ref{20}) follows from eq.~(\ref{9bis}), and eq.~(\ref{21})
follows
from
\be   \label{22}
\rho'+3\gamma \, \frac{a'}{a} \, \rho=0 \; ,
\ee
which can be derived from the conservation equation $\nabla^{\nu}
T_{\mu\nu}=0$ and the equation of state (\ref{11}). The scale factor
in the different eras is
\be
a( \eta) =\left\{
\begin{array}{cllll}
\label{23}
a_0^{(p)} \left[ \cos( \eta c_p) \right]^{1/c_p}
\;\;\;\;\;\;\;\;\;\;\;
0 \leq \eta \leq \eta_r  \; , \\
\label{24}
a_0^{(r)} \cos( \eta +d^{(r)} ) \;\;\;\;\;\;\;\;\;\;\;
\eta_r \leq \eta \leq \eta_m  \; , \\
\label{25}
a_0^{(m)}  \cos^2 ( \eta/2+d^{(m)})  \;\;\;\;\;\;\;\;\;
\eta_m \leq \eta \; .   \end{array}   \right.    \ee
By imposing that
the scale factor and its derivative (and hence, also the Hubble
parameter) are continuous at the transitions between the different
eras ($\eta_r$ and $\eta_m$), we derive appropriate values for the
integration constants:
\begin{eqnarray}
\label{26}  & & d^{(p)}=0 \; , \\
\label{27}  & & a_0^{(r)}=a_0^{(p)} \left[ \cos( \eta_r c_p)
\right]^{\frac{1}{c_p}-1} \; ,\\
\label{28}  & & d^{(r)}=\eta_r (c_p-1) \; , \\
\label{29}  & &   a_0^{(m)}=\frac{a_0^{(r)}}{ \cos( \eta_m +d^{(r)})} \; ,\\
\label{30}  & &   d^{(m)}=\frac{\eta_m}{2}+d^{(r)}  \; .
\end{eqnarray}

\subsection{Other quantities}

We will now derive expressions for the conformal times $\eta_r$ and
$\eta_m$ marking the duration of the prematter and radiation eras,
respectively. In our model, as in IR and SC, the universe starts in a very
cold state and heats during expansion in the inflationary era. This
follows from the equation of state and, as a result, there is no
requirement for ``re--heating'' as there is in other models of the
early universe. We assume, according to the above
mentioned idea of limiting values for the physical quantities, that
inflation stops when the temperature reaches the Planck value
$T_{pl}=1.4169 \cdot 10^{32}$~K. The energy density at this time
($\eta_r$) is
\be    \label{31}
\rho( \eta_r)=\frac{\pi^2}{15} \left( k T_{pl} \right)^4 \; ,
\ee
where $k=1.38 \cdot 10^{-16}$~K$^{-1}$ is the Boltzmann constant.

Equations (\ref{9bis}), (\ref{23}) and (\ref{31}), together with
(\ref{20}), (\ref{27}) and (\ref{28}), give
\be    \label{32}
\eta_r=\frac{1}{c_p} \arccos\left[ \left( \frac{15}{\pi^2} \,
\frac{\rho_{pl}}{(kT_{pl})^4} \right)^{c_p/3\gamma_p} \right] \; .  \ee
The dimensionless quantity in square brackets in eq.~(\ref{32}) will
be used in the following, and its numerical value, with the velocity of light
and the (reduced) Planck constant restored,
is
\be   \label{33}
\frac{15}{8\pi^5} \,\frac{\rho_{pl}}{(kT_{pl})^4}\,c^5 h^3=1.5201
\; .  \ee
We can derive an expression for $\eta_m$ by considering the evolution
of the temperature $T$, which is described by the equation
\be \label{34}
\gamma'-\gamma \, \frac{T'}{T} -3\,\frac{a'}{a} \,\gamma \, ( \gamma-1)=0
\ee
(corresponding to eq.~(2.20) in SC). The solution during
the radiation era ($\gamma=4/3$) is
\be   \label{35}
a( \eta) \, T( \eta)=\mbox{const.}=a( \eta_*) \, T( \eta_*) \; ,  \ee
where $\eta_*$ is any given value of $ \eta$ in the radiation era. In
particular, for $( \eta,\eta_*)=( \eta_m,\eta_r)$ we get, from Eqs.
(\ref{24}) and (\ref{28}),
\be    \label{36}
\eta_m=\arccos \left[ \frac{T_{pl}}{T_m} \,
\cos ( \eta_r c_p ) \right]+\eta_r (1-c_p) \; ,    \ee
where $T_m$ is the temperature at the last phase transition $\eta_m$.
Kolb and Turner (1990, p.~77) give
\be     \label{37}
T_m=T_{now} \cdot 2.32\cdot 10^4  \,\, \Omega_0 h^2 \;\; \mbox{K} \; ,
\ee
where $\Omega_0=\rho/\rho_c$, $\rho_c=3H^2/(8\pi)$ is the critical
density, $h=H_0/\left( 100 \; \mbox{Km}\cdot \mbox{s}^{-1}\cdot
\mbox{Mpc}^{-1} \right)$, and $T_{now}=2.7$~K. We use the numerical value
$\Omega_0 h^2=0.396$ (which is compatible with a rather wide range of
acceptable values for $\Omega_0$ and $h$), giving $ T_m=2.4805 \cdot
10^4$~K. Actually, it will turn out that our final results depend only
weakly on the value of $T_m$, and are much more strongly affected by
the choice of the parameter $\gamma_p$ in the prematter era.

We note that eq.~(\ref{36}) sets a limit on the acceptable values of
$\gamma_p$. Using eq.~(\ref{32}), the requirement that the argument of
the $\arccos$ function in eq. (\ref{36}) is not larger than 1 gives
\be    \label{38}
\gamma_p \leq \left\{ 3\left[ \frac{1}{2}
+\frac{\ln ( T_{pl}/T_m)}{\ln 1.5201}
\right] \right\}^{-1}   \; .
\ee
At this stage, it is useful to deduce the order of magnitude of the
quantities involved, and make some approximations. We first note that the
expression (\ref{23}) for the scale factor in the prematter era becomes
singular at $ \eta |c_p|=\pi/2$. Typical values of $\gamma_p$ that we
consider are of order $ 10^{-3}$, which gives
\be   \label{39}
N \equiv -\,\frac{|c_p|}{3\gamma_p} \simeq -300 \ee
and, from eq.~(\ref{32}),
\be    \label{40}
\eta_r |c_p|=\arccos (1.5201^N) \equiv \frac{\pi}{2}+\delta=
\frac{\pi}{2}-|\delta|  \; ,  \ee
with $ |\delta|\sim 10^{-55}$. Therefore, $\eta_r |c_p|<\pi/2$, and
inflation stops before the singularity in eq.~(\ref{23}) is reached, but
$\eta_r |c_p|$ is numerically very close to $\pi/2$.

We now introduce
\be            \label{41}
x \equiv \frac{T_{pl}}{T_m} \, |\delta|
\ee
which, for $T_m \sim 10^4$~K and $\gamma_p \sim10^{-3}$, is of order
$10^{-27}$. These numbers justify the following expansions: We have
\be   \label{42}
\cos ( \eta_r c_p )=\cos( \pi/2 +\delta)=-\sin \delta \simeq |\delta|
\ee
and eq.~(\ref{36}) gives
\be   \label{43}
\eta_m=\arccos x+\eta_r (1-c_p)=-\pi/2+x+\eta_r (1-c_p)+O(x^2)  \; ,
\ee
where we invert the function $\cos s $ in the interval $-\pi \leq s
\leq 0$. Any other branch of $\arccos x$ does not give
the desired (increasing) behavior of the scale factor after the last
phase transition.

We now compute the age of the universe, $\eta_{now}$, in
conformal time. To this end, we note that after the transition between
the radiation and matter eras, radiation is decoupled from matter, and
behaves like a perfect fluid with $\gamma=4/3$. Equation (\ref{35})
gives, for $( \eta,\eta_*)=( \eta_m,\eta_{now})$
\be
\label{44}
a( \eta_m) \, T_m=a( \eta_{now}) \, T_{now}
\ee
and using eq.~(\ref{25})
\be   \label{45}
\cos^2 \left( \frac{\eta_{now}}{2}+d^{(m)} \right)=\frac{T_m}{T_{now}}\,
\cos^2 \left( \frac{\eta_m}{2}+d^{(m)} \right)  \; .     \ee
Now, Eqs.~(\ref{43}), (\ref{28}) and (\ref{30}) give
\be   \label{46}
\frac{\eta_m}{2}+d^{(m)} \simeq \arccos x
\ee
and therefore
\be    \label{47}
\cos \left( \frac{\eta_{now}}{2}+d^{(m)}
\right)=\frac{T_{pl}}{\sqrt{T_m T_{now}}}\, |\delta|  \;  ,
\ee
where we have chosen the positive sign. We must now invert the
function $\cos s $ in the interval $( -\pi,0)$ as before. This branch
(which we call ``branch~2'') is the opposite of that obtained by inverting
$\cos s$ in the interval $(0,\pi)$ (denoted by ``branch~1''). Any branch
other than branch~2 will not give the desired behavior of the scale
factor (i.e. increasing for $\eta_m \leq \eta \leq \eta_{now}$, and
for $\eta_{now} \leq \eta \leq \eta_{max}$, where $\eta_{max}$ is the
time of maximal expansion of the universe before it starts
recollapsing). Therefore, we write
\be   \label{48}
\eta_{now}=-2 \left. \arccos\left[ \left( \frac{T_m}{2.7 \, K} \right)^{1/2}
\cdot x \right]\right|_{\mbox{{\small branch~1}}}-\eta_m-2d^{(r)}  \; .  \ee
Equation~(\ref{48}) gives an upper limit on $\gamma_p$. The
requirement that the argument of the $\arccos$ function in
eq.~(\ref{48}) is not larger than $1$ gives, using
eqs.~(\ref{39})--(\ref{42}),
\be  \label{newlimit}
\gamma_p \leq \frac{\ln 1.5201}{3\ln \left( \sqrt{1.5201}\, T_{pl}/\sqrt{T_m
T_{now}} \right)}   \; \; .                  \ee
The conformal time of maximum expansion of the universe is trivially
given by eq.~(\ref{25}) as
\be     \label{49}
\eta_{max}=-2d^{(m)}  \; ;     \ee
it is reached when
\be  \label{50}
a( \eta_{max})=a_0^{(m)}  \;  .
\ee

It is useful to translate conformal times into comoving times.
Conformal time is defined by eq. (\ref{trans}). Choosing $t=0$ at $\eta=0$
we get, from eq.~(\ref{17}),
\be \label{52}
t( \eta)=a_0 \int_0^{\eta}d\eta' \left[ \cos ( \eta' c+d)
\right]^{1/c} \; .
\ee
This integral has to be computed numerically in the prematter era,
while in the radiation ($c_r=1$) and matter ($c_m=1/2$) eras, the
integration is trivial. The expressions for the comoving times
corresponding to $\eta_r$, $\eta_m$, $\eta_{now}$ and $\eta_{max}$
are:
\be   \label{53}
t_r=a_0^{(p)} \int_0^{\eta_r}d\eta \, \frac{1}{\left[ \cos ( \eta c_p)
\right]^{1/|c_p|}} \; ,                   \ee
\be   \label{54}
t_m=t_r+a_0^{(r)} \left[ \sin ( \eta_m+d^{(r)})-\sin ( \eta_r c_p )
\right] \; ,                   \ee
\be   \label{55}
t_{now,max}=t_m+\frac{a_0^{(m)}}{2} \left[\eta_{now,max}-\eta_m+
\sin ( \eta_{now,max}+2d^{(m)})-\sin ( \eta_m+2d^{(m)} ) \right] \; .
\ee
The Hubble parameter at the present time is
\be \label{56}
H( \eta_{now})=\frac{a'( \eta_{now})}{a^2( \eta_{now})}=
-\,\frac{9.2503 \cdot 10^{29}}{a_0^{(m)}} \frac{
\sin ( \eta_{now}/2+d^{(m)})}{\cos^3 ( \eta_{now}/2+d^{(m)})} \;\;\;
\mbox{Km}\cdot \mbox{s}^{-1} \cdot \mbox{Mpc}^{-1}  .
\ee
The evolution of the energy density is governed by eq.~(\ref{22}),
which has the solution
\be   \label{57}
\rho \,a^{3\gamma}=\mbox{constant}=\rho( \eta_*) \, a^{3\gamma}( \eta_*)
\; ,    \ee
where $\eta_*$ is any fixed conformal instant during a given era. By
choosing $( \eta_*,\eta)=(0,\eta_r)$, $( \eta_r,\eta_m)$, $(
\eta_m,\eta_{now})$, $( \eta_m,\eta_{max})$ we get, respectively,
\begin{eqnarray}
\label{58} & & \rho( \eta_r)=\rho_{pl} \left[ \frac{a_0^{(p)}}{a( \eta_r)}
\right]^{3\gamma_p} \; , \\
\label{59}  & &  \rho( \eta_m)=\rho( \eta_r) \left[ \frac{a( \eta_r)}
{a( \eta_m)}\right]^4 \; ,     \\
\label{60} & &  \rho( \eta_{now})=\rho( \eta_m) \left[ \frac{a( \eta_m)}{a(
\eta_{now})} \right]^3 \; ,  \\
\label{61}  & &   \rho( \eta_{max})=\rho( \eta_m) \left[
\frac{a( \eta_m)}{a_0^{(m)}} \right]^3 \; .
\end{eqnarray}

\subsection{Numerical values}

We present tables with numerical values for the conformal and comoving
times marking the transitions between the prematter, radiation and
matter eras, for the present age of the universe and for the time of its
maximal expansion when it just starts recollapsing. We also give the
corresponding values for the scale factor (in cm) and energy density
(in g$\cdot$ cm$^{-3}$). The present value of the Hubble parameter
(in Km$\cdot$s$^{-1}$$\cdot$Mpc$^{-1}$) is also computed. The tables
correspond to the following choices of parameters: $T_m=2.4805 \cdot
10^4$~K and $\gamma_p=1.8500\cdot 10^{-3} $, $1.9000\cdot 10^{-3} $,
$1.9500\cdot 10^{-3} $, $2.0000\cdot 10^{-3} $, $2.0100\cdot 10^{-3} $,
$2.0153\cdot 10^{-3} $, which are allowed by the limits in eqs.~(\ref{38})
and (\ref{newlimit}).
It is to be noted that for a broad range of initial conditions
corresponding to values of $\gamma_p$ between $1.75 \cdot 10^{-3}$ and
$ 1.95 \cdot 10^{-3}$, the value
of the Hubble parameter hardly varies, remaining close to $46.7$
km$\cdot$~s$^{-1}\cdot$~Mpc$^{-1}$. If the preferred model of the
universe is one which does not demand the fine--tuning of parameters,
it could be argued that this is the proper value of $H_0$. While this
value is relatively low, it is compatible with the currently accepted
range.

It is to be noted that in the narrow range of allowed values of
$\gamma_p$ near the critical point from $2.01 \cdot 10^{-3} $ to $
2.0323 \cdot 10 ^{-3}$, the predicted
$H_0$ drops rapidly beyond the acceptable range. Moreover, while the
value of $H_0$ is substantively governed by the choice of the
parameter $\gamma_p$, it depends only weakly on the value of the
parameter $T_m$.

\section{The scalar field dynamics}

We turn now to the dynamics of the conformally coupled scalar field
driving inflation. As discussed above, we do not assume a priori
a form for the scalar field potential, but rather we derive it as a
consequence of the assumption that the equation of state changes in
the different eras of the universe. The Klein--Gordon equation
(\ref{2}) with $ \xi=1/6$ is\setcounter{equation}{0}
\be   \label{62}
\phi''+2\, \frac{a'}{a}\, \phi'+\left( \frac{a''}{a}+1 \right)
\phi+a^2\, \frac{dV}{d\phi}=0 \; .
\ee
Equations (\ref{11}), (\ref{7}) and (\ref{6}) give
\begin{eqnarray}
\gamma=\frac{P}{\rho}+1=2\left\{ \frac{2}{3}\left( \phi'
\right)^2+\frac{4}{3}\,\frac{a'}{a}\, \phi \, \phi'+\frac{1}{3} \,a^2 \phi
\, \frac{dV}{d\phi}+\frac{2}{3} \, \phi^2 \left[ 1+\left(
\frac{a'}{a}\right)^2 \right] \right\} \nonumber \\
\cdot \left\{ \left( \phi' \right)^2+2\, \frac{a'}{a} \, \phi \, \phi'+2a^2V+
\phi^2 \left[ 1+\left( \frac{a'}{a}\right)^2 \right] \right\}^{-1} \;
,  \label{63}
\end{eqnarray}
from which we get
\be   \label{64}
V=\frac{1}{2\gamma a^2}\left( \frac{4}{3}-\gamma \right) \left\{
\left( \phi' \right)^2+2\,\frac{a'}{a}\,\phi \, \phi'+\phi^2 \left[ 1+\left(
\frac{a'}{a}\right)^2 \right] \right\} +\frac{1}{3\gamma} \, \phi \,
\frac{dV}{d\phi}       \; .  \ee
Now, using eqs.~(\ref{6}) and (\ref{trans}) in eq.~(\ref{9}) we find
\be   \label{65}
V( \eta)=-\, \frac{1}{2a^2} \,( \phi')^2-\frac{3}{2a^2} \left(
\frac{\phi^2}{3}-\frac{1}{4\pi}\right) \left[ \left(
\frac{a'}{a}\right)^2+1 \right]-\frac{a'}{a^3}\, \phi \, \phi' \; .
\ee
Substituting the value of $dV/d\phi$ given by eq.~(\ref{62}) into
eq.~(\ref{64}), and the result into eq.~(\ref{65}), we find
\be   \label{66}
\phi \phi''-2\, \frac{a'}{a} \, \phi\phi'-2( \phi')^2+\phi^2\left[
\frac{a''}{a}-2\, \left( \frac{a'}{a}\right)^2-1\right] +
\frac{9\,\gamma}{8\pi}\left[ \left( \frac{a'}{a}\right)^2+1\right]
=0\; .    \ee
The last equation can be solved numerically in the prematter era, once
the parameter $\gamma_p$ is fixed, and initial conditions
$\phi_0$, $\phi'_0$ are specified. Then, one can solve numerically for
the values of $\phi'( \eta)$ and $V( \eta)$ from eq.~(\ref{65}) and,
by eliminating the conformal time, one obtains a numerical form for
the scalar field potential $V( \phi)$. Using the MAPLE computer
program, we have done this for $\gamma_p=2.0153 \cdot 10^{-3}$
(corresponding to Table~6). The result is plotted in two different
ranges of values of $\phi$. The resulting $V( \phi)$ does not depend
on the particular solution $\phi ( \eta)$ coming from specific initial
data $( \phi_0, \phi'_0)$, as must be the case. Integration gives only
narrow ranges of $\phi( \eta) $ for given initial conditions, so the
relation $\phi$~--~$V$ is plotted in narrow windows spanning only a part
of the possible range of values of $\phi$. We treat the initial
conditions as arbitrary. The question if there are preferred
initial conditions, and then a corresponding particular range for $\phi$
remains an open problem.

The features of the resulting $V( \phi)$ are as follows: the scalar
field potential has the symmetry $V( \phi)=V( -\phi)$, as to be
expected from an examination of eq.~(\ref{65}), and for $\phi>0$ the
overall shape is plotted in fig.~1. $V$ is slightly positive for small values
of $\phi$, and becomes negative as $\phi$ increases. Superposed upon
the basic overall shape are variations of $V( \phi)$ on a small scale, which
occur in narrow ranges of $\phi$; some of these variations, with local maxima
and minima are plotted in fig.~2.

If one accepts the above mentioned idea that the scalar field is the dominant
component of dark matter also in the post--inflationary eras of the universe,
it is immediate to derive the form of the potential $V$ in the radiation
era. In fact, the trace of the energy--momentum tensor is
\be  \label{trace}
{T^{\alpha}}_{\alpha}=\rho-3P=4V-\phi \,\frac{dV}{d\phi} \; ,
\ee
which vanishes when $ \gamma=4/3$, giving
\be  \label{quarticpotential}
V( \phi)=\lambda \phi^4  \; ,
\ee
where $\lambda$ is a constant. This form of the scalar field potential
has been used widely in the literature. It should be noted that this
result holds only in the case of conformal coupling. For example, when
$\xi=0$, SC obtained a different form of $V $.

\section{Discussion and conclusion}

We have constructed a closed cosmological model which has no
singularities. In contrast to other works (e.~g. Balbinot {\em et al.} 1990,
Brandenberger {\em et al.} 1993), we use only general relativity, and an
inflationary equation of state in the early universe. We impose
discontinuous changes in the equation of state, and solve for the
dynamics of the universe. The key ingredient to obtain a
singularity--free model is the Gliner--Markov idea that the physical
quantities are limited by the Planck values. Our boundary and initial
conditions on the scale factor reflect this idea. The
evolution equations for the scale factor and the energy density can be
solved exactly by using conformal time instead of the more commonly
used comoving time. However, the physically significant conformal
times have been converted to comoving times. We considered three
different eras for the history of the universe, corresponding to
different values of the constant $\gamma$ in the equation of state.
The early inflationary era is dominated by prematter, then a
radiation--dominated and a matter--dominated era follow. We determined
the times of transition between the different eras, the present age of
the universe, the time of its maximal expansion, and the Hubble
parameter, for various values of the parameter $\gamma_p$. We obtained
values of $H_0$ and $t_{now}$ in a range which agrees with the
observations and the current theoretical expectations.

The only source of gravitation is provided by a scalar field, which is
chosen to be conformally coupled to the Ricci curvature, for the above
mentioned reasons. We proceeded to the numerical solution of the
equations for the scalar field and the potential as functions of
conformal time. From these, we derived the form of the potential $V(
\phi)$. It is interesting to note that the derived potential differs
from that computed in SC for a minimally coupled field.
Moreover, the greater complexity of the non--minimal coupling prevents one
from deriving a differential equation, and then an analytical expression
for $V( \phi)$, as was possible in the SC paper where the coupling was
minimal.

The potential $V( \phi)$ exhibits variations on a small scale, with local
maxima and minima. One could ask which range of values of $\phi$ is
relevant to the physics of the early universe, and hence what is the
shape of the potential for the appropriate range of the scalar field. The
answer to this problem is not an easy one, since the actual values of
$\phi$ are determined by the arbitrary initial data set $(
\phi_0,\phi'_0)$. Although one can try to constrain the value of $\phi_0$
with arguments based on mass scales (a very speculative approach, but
one which has received some attention in the literature~--~see e.g.
Hosotani 1985 and Futamase \& Maeda 1989),
there are no indications on how to constrain the initial derivative
$\phi'_0$. It turns out that the minimally coupled case
requires only the consideration of one initial value, $\phi_0$.

It must be noted that various inflationary scenarios with a
nonminimally coupled scalar field and an {\em a priori} chosen form of
the potential encounter problems (Abbott 1981; Starobinsky 1981;
Futamase \& Maeda 1989; Futamase {\em et al.} 1989; Bruni 1993). On
the other hand, conformal coupling is required if the Einstein
equivalence principle is to hold. In our approach, because of the fact
that we do not fix a form of the potential, but rather derive it, the
previous difficulties are avoided, and we can have both conformal
coupling and a viable inflation (at least to the extent explored in
this paper).

Our model universe starts very cold and initially static from Planck
size and density, and undergoes an inflationary expansion, during
which it heats. It must be noted that the possibility of imposing the
initial condition $ a'(0)=0$, and the absence of an initial singularity
are by no means trivial, in the light of the Hawking--Penrose singularity
theorems (see the formulations in Hawking \& Ellis 1973; Wald 1984;
and references therein). In fact, the strong energy
condition required by these theorems, which can be written
as\setcounter{equation}{0}
\be \label{67}
\rho+P\geq 0\;\;\;\;\;\mbox{and}\;\;\;\;\rho+3P\geq 0 \; ,
\ee
is violated for $\gamma< 2/3$, which is our case, since $\gamma_p \sim
10^{-3}$. On the contrary, in
the standard model starting from the big
bang in a radiation--dominated epoch, $\gamma=4/3$. The weak ($\rho \geq 0$
and $\rho+P \geq0$) and dominant ($\rho\geq |P|$) energy conditions are
satisfied. The inflationary equation of state permits the avoidance
of singularities, and in general relativity, inflation (defined as positive
acceleration of the scale factor) occurs if and only if the
condition $\rho+3P\geq 0$ is violated.

Regarding the regularity of the functions of physical interest,
the continuity of the scale factor $a( \eta)$ and its derivative were
assured by the proper choice of integration constants in eqs.
(\ref{26})--(\ref{30}). Thus, the Lichnerowicz (1955) conditions on the
metric are satisfied and from eq.~(\ref{9}), the density $ \rho$ is
continuous at the phase transitions. Since we choose different constant
values of $ \gamma$ at the three different universe epochs ($\gamma_p
\sim 10^{-3}$, $\gamma_r=4/3$, $\gamma_m=1$ respectively), the pressure
from eq.~(\ref{11}) is discontinuous at the phase
transitions.

Since the idea of treating the scalar field as the primary source of
gravitation in the radiation-- and matter--dominated epochs is a
speculation at this stage, we only found $V( \phi)$ in the radiation era
(eq.~(\ref{quarticpotential})) and we did not present a numerical solution
for $ \phi$ and $V( \phi)$ otherwise for those epochs. We have treated
the initial data $( \phi_0, \phi'_0)$ for the scalar field as arbitrary and
we can use the values of $\phi$ and $\phi'$ at the end of a given epoch as
the initial values for the succeeding epoch. Thus, it is always possible
to evolve a continuous solution in $\phi$, $\phi'$. However, since the
potential $V( \phi)$ has a different functional form in the different
epochs, corresponding to the different equations of state, it is not
continuous. The difference in form is evident in comparing the numerical
solution in figs.~1,~2 with the analytical form in
eq.~(\ref{quarticpotential}). However, if $\phi( \eta)$, $\phi'(
\eta)$ are continuous, then the potential as a function of the conformal
time, $V( \eta)$, will be continuous as can be seen in eq.~(\ref{6}).
However, this is not inconsistent with the discontinuity of $V$ as expressed
as a function of $\phi$.

After the universe reaches the point of maximal expansion, it
experiences its past history again in reversed order, until it reaches
a point, in a prematter era, in which the derivative of the scale
factor vanishes, and then the past history is repeated. The vanishing
of $a'$ makes it possible to match the two FLRW metrics before and after
the bounce in such a way that the scale factor and its derivative are
continuous, thus satisfying the Lichnerowicz (1955) conditions.

We did not address the problem of density perturbations and the
formation of structures in our model universe. Perturbations of a
non--minimally coupled scalar field suffer from the problems pointed
out in sec.~2, and will be the subject of future analysis. Possible
generalizations of this paper include the study of open models and of
inhomogeneous metrics.

\section*{Acknowledgments}

The assistance of Philip Perry with numerical computations is gratefully
acknowledged. This research was supported, in part, by a grant from the
Natural Sciences and Engineering Research Council of Canada. This work
was partially supported by TUBITAK, the Scientific and Research
Council of Turkey under TBAG/CG--1.

\vspace*{1truecm}  {\small
\noindent {\bf References} \\

\noindent Abbott, L. F. 1981, Nucl. Phys. B 185, 233

\noindent Balbinot, R., Fabris, J. C. \& Kerner, R. 1990, Class. Quantum
Grav. 7, L17

\noindent Barrow, J. 1990, Phys. Lett. B 235, 40

\noindent Birrell, N. D. \& Davies, P. C. 1982, Quantum Fields in Curved
Space (Cambridge: Cambridge Univ. Press)

\noindent Bruni, M. 1993, private communication

\noindent Brandenberger, R., Mukhanov, V. \& Sornborger, A. 1993,
 Phys. Rev. D 48, 1629

\noindent Copeland, E. J., Kolb, E. W., Liddle, A. R. \& Lidsey, J. E.
1993, Phys. Rev. D 48, 2529; Phys. Rev. Lett. 71, 219

\noindent Delgado, V. 1993, ULLFT--1/93 preprint

\noindent Ellis, G. F. R. 1991, in Gravitation, Banff Summer Institute,
August 12--25 1990, ed. R. Mann \& P. Wesson (Singapore: World
Scientific), p.~3

\noindent Ellis, G. F. R. \& Madsen, M. S. 1991, Class. Quantum Grav. 8, 667

\noindent Futamase, T. \& Maeda, K. 1989, Phys. Rev. D39, 399

\noindent Futamase, T., Rothman, T. \& Matzner, R. 1989, Phys. Rev. D 39,
405

\noindent Gliner, E. B. 1966, Sov. Phys. JETP 22, 378

\noindent Hawking, S. W. \& Ellis, G. F. R. 1973, The Large Scale Structure
of Space--Time (Cambridge: Cambridge Univ. Press)

\noindent Hodges, H. M. \& Blumenthal, G. R. 1990, Phys. Rev. D 42, 3329

\noindent Hosotani, Y. 1985, Phys. Rev. D 32, 1949

\noindent Israelit, M. \& Rosen, N. 1989, ApJ 342, 627 (IR)

\noindent Kofman, L., Linde, A. \& Starobinsky, A. in Proceedings of the
5th Canadian Conference on General Relativity and Relativistic Astrophysics
(World Scientific), to appear

\noindent Kolb, E. \& Turner, M. 1990, The Early Universe (New York:
Addison--Wesley)

\noindent Lichnerowicz, A. 1955, Th\'{e}ories Relativistes de la
Gravitation et de l' Electromagnetisme (Paris: Masson)

\noindent Lidsey, J. E. \& Tavakol, R. K. 1993, Phys. Lett. B 309, 23

\noindent Lucchin, F. \& Matarrese, S. 1985, Phys. Rev. D 32, 1316

\noindent Madsen, M. S. 1988, Class. Quantum Grav. 5, 627

\noindent Markov, M. A. 1982, Sov. Phys. JETP Lett. 36, 265

\noindent McDonald, J. 1993, Phys. Rev. D 48, 2462

\noindent Rosen, N. 1985, ApJ 297, 347

\noindent Sonego, S. \& Faraoni, V. 1993, Class. Quantum Grav. 10,
1185

\noindent Starkovich, S. P.  1992, PhD thesis, University of Victoria

\noindent Starkovich, S. P.  \& Cooperstock, F. I. 1992, ApJ 398, 1
(SC)

\noindent Starobinsky, A. A. 1981, Sov. Astron. Lett. 7, 36

\noindent Synge, J. L. 1955, Relativity: the General Theory (Amsterdam: North
Holland)

\noindent Turner, M. S. 1993, FERMILAB--PUB--93/182--A preprint

\noindent Wald, R. M. 1984, General Relativity (Chicago: The Univ. of
Chicago Press)   }

\vspace*{1truecm}
\noindent {\bf Table and figure captions:} \\ \\
{\bf Table~1:} results for $\gamma_p=1.8500 \cdot 10^{-3}   $. \\ \\
{\bf Table~2:} results for $\gamma_p=1.9000 \cdot 10^{-3}   $. \\ \\
{\bf Table~3:} results for $\gamma_p=1.9500 \cdot 10^{-3}   $. \\ \\
{\bf Table~4:} results for $\gamma_p=2.0000 \cdot 10^{-3}   $. \\ \\
{\bf Table~5:} results for $\gamma_p=2.0100 \cdot 10^{-3}   $. \\ \\
{\bf Table~6:} results for $\gamma_p=2.0153 \cdot 10^{-3}   $. \\ \\
{\bf Figure~1:} the overall shape of the scalar field potential $V(
\phi)$ (rescaled by $10^{68}$ cm$^{-2}$, in units such that $G=c=1$) for
$\phi>0$. The plot is
obtained by superposing different windows like that in fig.~2, each
spanning a small range of values of $\phi$. The plots in fig.~1 and
Fig.~2 correspond to $T_m=2.4805 \cdot 10^4 $~K and $\gamma_p=2.0153
\cdot 10^{-3}$.  \\ \\
{\bf Figure~2:} an example of the variations of $V( \phi)$ on a small
scale, with local maxima and minima.
\end{document}